\pdfoutput=1 
\documentclass[]{spie}  %>>> use for US letter paper
\usepackage[]{graphicx}
\usepackage{amssymb, amsmath}

\newcommand{\lya}{Ly$\alpha$}
\newcommand{\arcmin}{$^{\prime}$}
\newcommand{\arcsec}{$^{\prime\prime}$}
\newcommand{\degree}{$^{\circ}$}
\title{Mass production of volume phase holographic gratings for the VIRUS spectrograph array} 

\author{Taylor S. Chonis\supit{a}, Amy Frantz\supit{b}, Gary J. Hill\supit{a,c},  J. Christopher Clemens\supit{b,d}, Hanshin Lee\supit{c}, Sarah E. Tuttle\supit{c}, Joshua J. Adams\supit{e}, J.L. Marshall\supit{f}, D.L. DePoy\supit{f}, Travis Prochaska\supit{f}
\skiplinehalf
\supit{a}The University of Texas at Austin, Department of Astronomy, 2515 Speedway, Stop C1400, Austin, TX, USA 78712; \\
\supit{b}Syzygy Optics, LLC, P.O. Box 211, Chapel Hill, NC, USA 27514; \\
\supit{c}The University of Texas at Austin, McDonald Observatory, 2515 Speedway, Stop C1402, Austin, TX, USA 78712; \\
\supit{d}University of North Carolina, Department of Physics \& Astronomy, CB3255, Chapel Hill, NC, USA 27599; \\
\supit{e}Observatories of the Carnegie Institution for Science, 813 Santa Barbara St., Pasadena, CA, USA 91101;\\
\supit{f}Texas A\&M University, Department of Physics \& Astronomy, 4242 TAMU, College Station, TX, USA 77843
}

\authorinfo{Further author information: (Send correspondence to T.S.C.)\\T.S.C.: E-mail: tschonis@astro.as.utexas.edu, Telephone: 1-512-471-1495}

\begin{document} 
  \maketitle 
%%%%%%%%%%%%%%%%%%%%%%%%%%%%%%%%%%%%%%%%%%%%%%%%%%%%%%%%%%%%% 

\begin{abstract}
The Visible Integral-field Replicable Unit Spectrograph (VIRUS) is a baseline array of 150 copies of a simple, fiber-fed integral field spectrograph that will be deployed on the Hobby-Eberly Telescope (HET). VIRUS is the first optical astronomical instrument to be replicated on an industrial scale, and represents a relatively inexpensive solution for carrying out large-area spectroscopic surveys, such as the HET Dark Energy Experiment (HETDEX). Each spectrograph contains a volume phase holographic (VPH) grating with a 138 mm diameter clear aperture as its dispersing element. The instrument utilizes the grating in first-order for $350<\lambda\;\mathrm{(nm)}<550$. Including witness samples, a suite of 170 VPH gratings has been mass produced for VIRUS. Here, we present the design of the VIRUS VPH gratings and a discussion of their mass production. We additionally present the design and functionality of a custom apparatus that has been used to rapidly test the first-order diffraction efficiency of the gratings for various discrete wavelengths within the VIRUS spectral range. This device has been used to perform both in-situ tests to monitor the effects of adjustments to the production prescription as well as to carry out the final acceptance tests of the gratings' diffraction efficiency. Finally, we present the as-built performance results for the entire suite of VPH gratings.
\end{abstract}

%>>>> Include a list of keywords after the abstract 
\keywords{Volume phase holographic diffraction gratings, Gratings: performance, Gratings: testing, Gratings: production, Spectrographs: integral field, VIRUS, HETDEX, Hobby-Eberly Telescope}
%%%%%%%%%%%%%%%%%%%%%%%%%%%%%%%%%%%%%%%%%%%%%%%%%%%%%%%%%%%%%

%%%%%%%%%%%%%%%%%%%%%%%%%%%%%%%%%%%%%%%%%%%%%%%%%%%%%%%%%%%%%
\section{INTRODUCTION} \label{sec:intro}  % \label{} allows reference (\ref{}) to this section
\begin{figure}[t]
   \begin{center}
   \begin{tabular}{c}
   \includegraphics[width=0.95\textwidth]{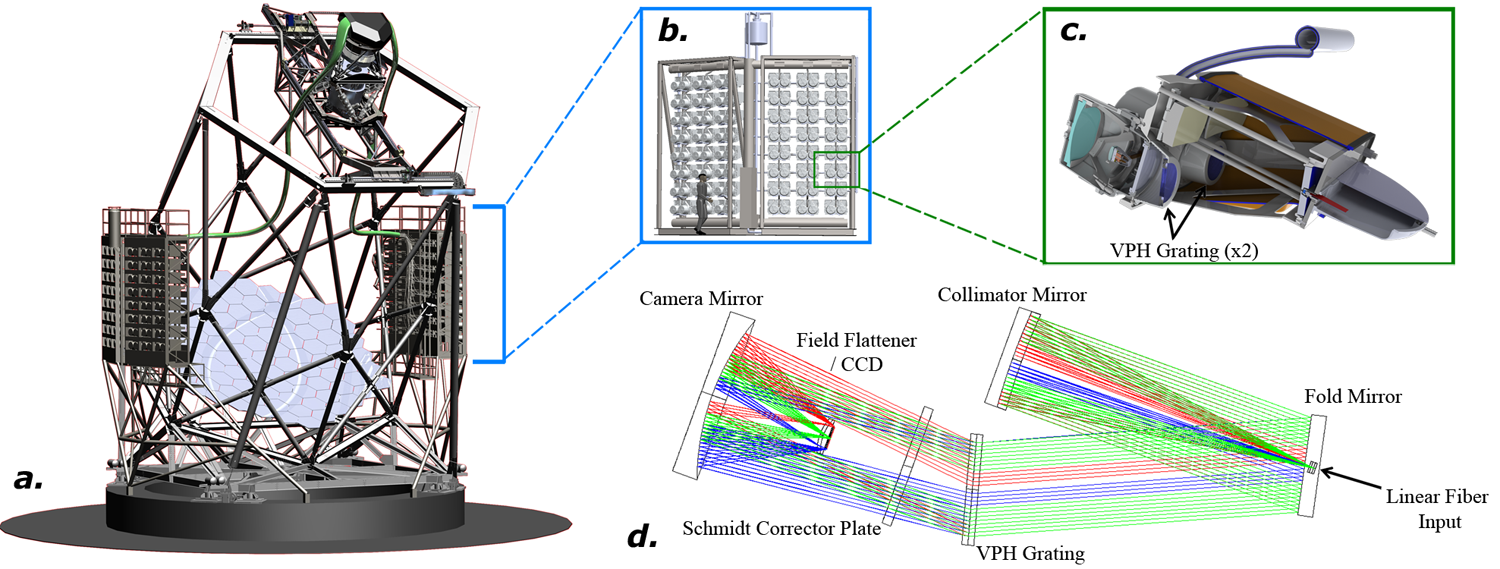}
   \end{tabular}
   \end{center}
   \caption[example] 
   { \label{fig:VIRUS} 
\textit{a}) A rendering of the upgraded HET showing the large enclosures mounted on either side of the telescope structure that contain VIRUS spectrographs. The green cables extending from the prime-focus instrument package at the top of the telescope structure to the enclosures are large bundles of fiber optics. \textit{b}) Close view of two enclosures, each containing an 8$\times$3 array of VIRUS units (48 spectrographs). \textit{c}) Section view of a single VIRUS pair. \textit{d}) Ray trace for a single VIRUS spectrograph.}
   \end{figure}

The upcoming Hobby-Eberly Telescope Dark Energy eXperiment (HETDEX; Ref. \citenum{Hill08a}) will amass a sample of $\sim0.8$ million \lya\ emitting galaxies (LAEs) that will be used as tracers of large-scale structure for constraining dark energy and measuring its evolution from $1.9 < z < 3.5$. To carry out the 120 night blind spectroscopic survey covering a 420 square degree field (9 Gpc$^{3}$), a revolutionary new multiplexed instrument called VIRUS (the Visible Integral field Replicable Unit Spectrograph; Ref. \citenum{Hill14a}) is being constructed\cite{Tuttle14} for the upgraded 9.2 m Hobby-Eberly Telescope (HET\footnote{The Hobby-Eberly Telescope is operated by McDonald Observatory on behalf of the University of Texas at Austin, the Pennsylvania State University, Ludwig-Maximillians-Universit\"{a}t M\"{u}nchen, and Georg-August-Universit\"{a}t G\"{o}ttingen.}; Ref. \citenum{Hill14b}). The VIRUS array consists of at least 150 copies of a simple fiber-fed integral field spectrograph and is the first optical astronomical instrument to leverage the economies of scale associated with large-scale replication to significantly reduce overall costs. The spectrographs are mechanically built into unit pairs and are fed by dense-pack fiber bundle integral field units (IFU) with 1/3 fill factor, each consisting of 448 fiber optic elements with a core diameter of 266 $\mu$m (1.5\arcsec\ on the sky). Thus, each individual spectrograph images 224 fibers. At least 75 IFUs will be arrayed on the 22\arcmin\ diameter focal plane of the upgraded HET, yielding $\sim33,000$ individual spectra per exposure. Each spectrograph consists of a double-Schmidt optical design with a volume phase holographic (VPH) diffraction grating at the pupil between a $f$/3.33 folded collimator and a $f$/1.25 cryogenic camera. The spectral coverage of VIRUS is $350 < \lambda \mathrm{(nm)} < 550$ at $R = \lambda / \Delta\lambda \approx 700$, which is optimal for measuring the baryonic acoustic oscillation via the \lya\ emission of star-forming galaxies from $1.9 < z < 3.5$. Fig. \ref{fig:VIRUS} shows a rendering of VIRUS and its optical design. 

%-------------
%------------- 

The VIRUS concept was proven by the Mitchell Spectrograph (formerly known as VIRUS-P; Ref. \citenum{Hill08b}), a single prototype VIRUS spectrograph that has been in use at the McDonald Observatory 2.7 m Harlan J. Smith telescope since 2007. The instrument has excellent throughput in the blue down to 350 nm ($\sim30$\%, excluding the telescope and atmosphere). For VIRUS, better throughput is required to maximize the number of LAE detections in order to achieve the goals of HETDEX. This is especially true at the lowest redshifts of the survey (i.e., at wavelengths in the near-ultraviolet) because the surveyed volume is smaller, the number density of bright LAEs is diminished (e.g., Ref. \citenum{Ciardullo12}), and the atmospheric transmission is quickly decreasing.

An optical component that can be improved in efficiency at these wavelengths is the VPH diffraction grating that is used as the instrument's dispersing element. VPH gratings have become the standard in astronomical spectroscopy as they provide higher diffraction efficiency and versatility over classic surface relief gratings\cite{Barden98}. For an overview of the physics of VPH gratings, we refer the reader to Refs. \citenum{Arns99}, \citenum{Barden00}, and \citenum{Baldry04}. Ref. \citenum{Adams08} discusses the performance of VPH gratings developed for the Mitchell Spectrograph, which at that time pushed the technology to the highest diffraction efficiency achieved at 350 nm ($\sim60$\%). More recently, as shown in Ref. \citenum{Chonis12}, we have developed prototype VPH gratings for VIRUS that have achieved diffraction efficiencies of $\gtrsim70$\% at 350 nm. For VIRUS, a key technological challenge will be to achieve consistency in this high performance standard over a large production suite of 170 gratings. 

In this paper, we present the mass production of VPH gratings for VIRUS with a focus on the acceptance testing methodologies and the as-built performance of the suite of 170 gratings. We begin in $\S$\ref{sec:gratingspec} by describing the production design of the gratings. In $\S$\ref{sec:evaluation}, we discuss the performance metrics by which the gratings are judged, and present the design of a custom apparatus that ensures the standardization of our acceptance tests. In $\S$\ref{sec:production}, the grating mass production at Sygygy Optics, LLC is discussed and the resulting performance of the grating suite is presented in $\S$\ref{sec:performance}. Our final thoughts and a summary are provided in $\S$\ref{sec:conclusions}.
%%%%%%%%%%%%%%%%%%%%%%%%%%%%%%%%%%%%%%%%%%%%%%%%%%%%%%%%%%%%%

%%%%%%%%%%%%%%%%%%%%%%%%%%%%%%%%%%%%%%%%%%%%%%%%%%%%
\section{VIRUS VPH GRATING PRODUCTION DESIGN} \label{sec:gratingspec}
%-------------
   \begin{figure}[t!]
   \begin{center}
   \begin{tabular}{c}
   \includegraphics[width=0.92\textwidth]{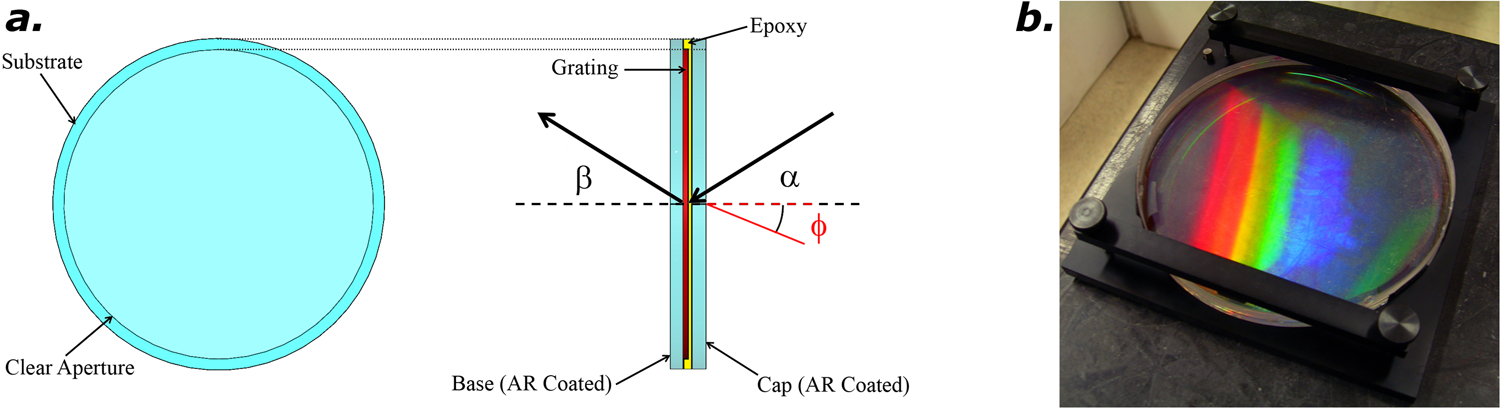}
   \end{tabular}
   \end{center}
   \caption[example] 
   { \label{fig:GratingSchem} 
\textit{a}) A schematic drawing of the VIRUS VPH diffraction grating, showing a face-on view of the grating and an edge-on view. The substrate diameter is 148 mm and the total thickness of the grating assembly is 16 mm. Note that the grating and epoxy layer thicknesses are exaggerated and are not to scale. The incident angle $\alpha$ and angle of diffraction $\beta$ are shown in addition to the direction of the fringe tilt, indicated in red by $\phi$. \textit{b}) A photograph of a production VIRUS VPH grating.\vspace{5mm}
}
   \end{figure} 
%------------- 
\begin{table}[t]
\caption{VIRUS external diffraction efficiency requirement for unpolarized light. All values are assumed to be averaged over the grating's clear aperture.} 
\label{tab:Efficiency}
\begin{center}       
\begin{tabular}{|c|c|c|} 
\hline
\rule[-1ex]{0pt}{3.5ex}  \textbf{$\lambda$} & \textbf{Batch Mean} & \textbf{Batch Minimum} \\
\rule[-2ex]{0pt}{1ex}  \textbf{[nm]} & \textbf{[\%]} & \textbf{[\%]} \\
\hline
\rule[-1ex]{0pt}{3.5ex}  350 & 70 & 60 \\
\hline
\rule[-1ex]{0pt}{3.5ex}  450 & 70 & 60 \\
\hline
\rule[-1ex]{0pt}{3.5ex}  550 & 40 & 30 \\
\hline
\end{tabular}
\end{center}
\end{table}

A schematic drawing and photograph of a VIRUS production grating can be seen in Fig. \ref{fig:GratingSchem}. The grating assembly has physical dimensions of 148 mm (diameter) $\times$ 16 mm (total thickness). The VPH layer has a 138 mm diameter clear aperture (CA) and is sandwiched between two 8 mm thick, anti-reflection (AR) coated fused silica substrates using an optical grade adhesive. The grating has a fringe density of $930\pm2$ lines mm$^{-1}$, which provides a level of dispersion that is sufficient to cover $350<\lambda\;\mathrm{(nm)}<550$ at spectral order $m=1$. The grating will operate in transmission for unpolarized light. The key properties of the gratings are high diffraction efficiency (especially for the bluer wavelengths) and repeatability of the grating properties from unit to unit. Due to the large number of units required for VIRUS (170 science-grade gratings, plus four witness samples of lesser quality to monitor environmental degradation over the lifetime of the gratings), the gratings were delivered in batch sizes of up to 50 units (but no smaller than 10) over a 12 month time period. To accommodate the large number of units and the expected variation of performance from unit to unit, the required external diffraction efficiency for unpolarized light was defined as a mean over a given delivery batch. Uniformity from unit to unit is promoted by establishing a minimum allowable efficiency for any grating. The batch mean and minimum external diffraction efficiencies are summarized in Table \ref{tab:Efficiency}. 

For HETDEX, the ideal peak diffraction efficiency is between $350 < \lambda\; \mathrm{(nm)} < 400$, and we have focused on maximizing the diffraction efficiency over this wavelength range while maintaining a sufficiently wide bandwidth to retain acceptable efficiency towards 550 nm. Given the parameters listed in this section, the Bragg condition (e.g., see Ref. \citenum{Baldry04}) dictates that the grating angle of incidence $\alpha$ is $\sim10$\degree\ to maximize the diffraction efficiency over the desired wavelength range. However, Ref. \citenum{Burgh07} has shown that the location in detector space where the wavelength satisfying the Bragg condition is imaged is also the location of the ``Littrow recombination ghost''. This optical ghost can have a wavelength integrated strength that dominates the signal in a given resolution element of the direct spectrum and can masquerade as a solitary emission line source\cite{Adams08}. For HETDEX, this could contribute significantly to sample contamination since normal LAE detections in our redshift range do not include any other bright emission lines other than \lya\ itself. To mitigate this issue, the fringes are tilted by $\phi = -1$\degree\ to decouple the Bragg condition from the Littrow configuration. Note that we have adopted the sign convention of Ref. \citenum{Burgh07} for $\phi$, where negative tilts move the plane of the fringes away from the incident beam, as depicted in Fig. \ref{fig:GratingSchem}$a$. By including $\phi$, we reduce the angle of incidence on the grating substrate to 9\degree. This allows the retention of the diffraction efficiency curve similar to a $\alpha=10$\degree\ grating with unslanted fringes while pushing the Littrow ghost off the CCD detector as a result of the change in the physical grating angle.
%%%%%%%%%%%%%%%%%%%%%%%%%%%%%%%%%%%%%%%%%%%%%%%%%%%%

%%%%%%%%%%%%%%%%%%%%%%%%%%%%%%%%%%%%%%%%%%%%%%%%%%%%
\section{EVALUATING THE VPH GRATING SUITE} \label{sec:evaluation}

\subsection{Evaluation of Prototype Gratings}\label{subsec:prototypegratings}
Prior to engaging in the full-scale production of the VIRUS VPH gratings as described in $\S$\ref{sec:production}, we carried out multiple design studies with several vendors to tune the VPH grating prescription sufficiently to meet the requirements of the HETDEX survey. These efforts have been documented extensively in Ref. \citenum{Adams08} for the Mitchell Spectrograph and Ref. \citenum{Chonis12} for VIRUS. Both of these publications include tests carried out with a custom automated test facility that allowed the full characterization of a grating over a range of $\alpha$, angle of diffraction $\beta$, and $m$. This test-bench was essential in the determination of the final specifications to which the VIRUS production gratings were built. 

\subsection{Acceptance Test Requirements} \label{subsec:testrequirements}
Efficiently and consistently testing 170 gratings to validate acceptance metrics requires a different approach than the detailed characterization efforts described in Refs. \citenum{Adams08} and \citenum{Chonis12}. The following is a brief list of requirements for the acceptance tests we have performed on the mass-produced gratings:

\begin{itemize} \itemsep1pt \parskip0pt \parsep0pt
	\item The testing method must provide a standardized reference for comparison to specifications.
	\item Characterization and testing must take no longer than 10 minutes per grating.
	\item Diffraction efficiency measurements at $\alpha=9$\degree\ and $m=1$ must be made for $\geq3$ wavelengths within $350 < \lambda \mathrm{(nm)} < 550$.
	\item Tests must provide an estimation of the spatial uniformity of the diffraction efficiency across the CA.
	\item Tests must provide an estimation of the scattered light in the VPH layer for at least one wavelength in the near-ultraviolet (UV).
	\item The test apparatus must be transportable to the vendor's facility and be operable in an office environment.
	\item The test apparatus' calibration and data reduction must be transparent to the operator.	
\end{itemize}

The test-bench discussed in the previous subsection is too large to be easily transported, and a test of a grating for multiple subapertures to provide a measure of the spatial uniformity is time consuming. Additionally, the ability to test a grating for a range of $\alpha$, $\beta$, and $m$ is not necessary for the acceptance tests. More precisely, the flexibility is undesirable as it may increase the probability of user error over a large number of grating tests. As we describe in the following subsection, we have designed a new apparatus to meet these requirements and efficiently check-out a grating directly on the production line.

\subsection{Design and Operation of the VIRUS Grating Tester}\label{subsec:testerdesign}
The design, operation, and validation of the prototype of the new grating test apparatus was described in Ref. \citenum{Chonis12}. We constructed a second, more refined copy of the apparatus and sent it to Syzygy Optics, LLC to be integrated into the production line they setup for the VIRUS VPH grating contract. Hereafter, we refer to this device as the VIRUS Grating Tester (VGT). The original prototype apparatus has remained at the University of Texas at Austin and has been used to check the results from the VGT and to characterize witness samples. 

%-------------
   \begin{figure}[t]
   \begin{center}
   \begin{tabular}{c}
   \includegraphics[width=0.8\textwidth]{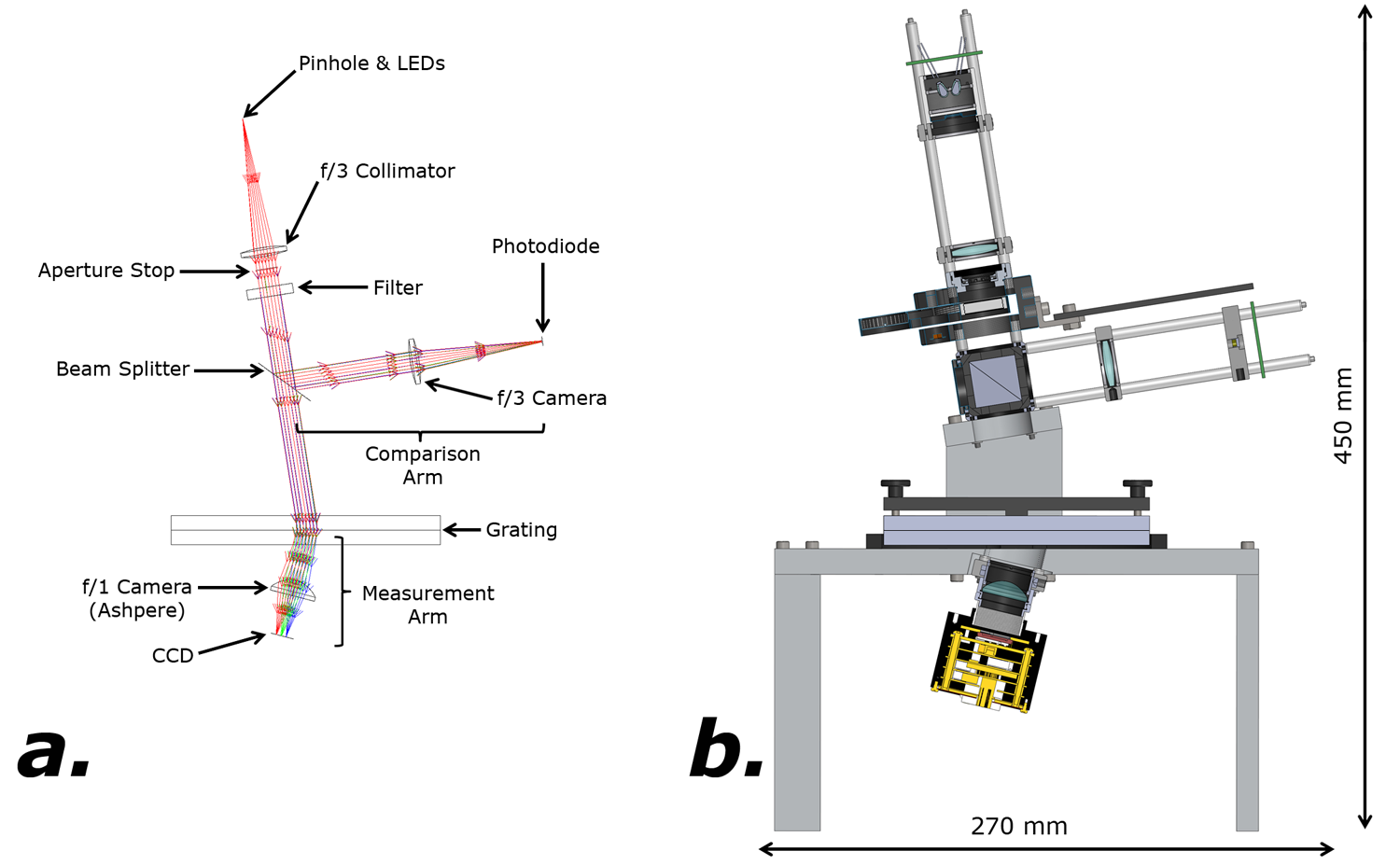}
   \end{tabular}
   \end{center}
   \caption[example] 
   { \label{fig:design} 
The optomechanical design of the VGT. $a$) Ray trace of the VGT optics with the major components labeled. $b$) A cross-section of the VGT mechanical model shown at the same scale and orientation as the ray trace in panel $a$ with rough dimensions indicated for scale.
}
   \end{figure} 
%------------- 

Fig. \ref{fig:design} shows the optomechanical design of the VGT. For direct comparison to the external diffraction efficiency specification in Table \ref{tab:Efficiency}, the VGT performs measurements only at 350, 450, and 550 nm. The light from the LED sources first passes through an engineered diffuser, followed by a 300 $\mu$m diameter pinhole that is placed at the focus of a 25 mm diameter $f$/3 singlet. The collimated beam is then stopped down to 12.5 mm in diameter. The 12.5 mm beam size was chosen to avoid the differential vignetting of the dispersed collimated beam after transmission through the grating since the collimator and camera lenses have the same physical diameter (see below). With an emitted FWHM of approximately 25, 25, and 60 nm for the 350, 450 and 550 nm LEDs, respectively, the light sources are far from monochromatic. To centralize the spectral emission, we limit each respective LED by using a 10 nm FWHM narrow-band filter in the collimated beam. The final, effective measurement wavelengths after filtering the LEDs' output are 353.9, 452.3, and 549.5 nm.

After the filter, the collimated beam is split into two paths. The first path is refocused with a second $f$/3 singlet onto a silicon photodiode, which is used to monitor and self-calibrate instabilities in the LED output. The active area of the photodiode is 1.6 mm in diameter, which is large enough to accommodate the chromatic aberration attributed to the singlet lenses. The second path of the collimated beam is incident on the diffraction grating at $\alpha = 9$\degree. The diffracted light ($\beta = 15.3$\degree\ at 450 nm) is then focused onto a 2/3''-format, 5 megapixel CCD (3.45 $\mu$m square pixels) by a stock 25 mm diameter $f$/1 aspheric lens.  The use of a CCD (rather than multiple single pixel detectors for each wavelength, such as photodiodes) is highly beneficial for the VGT. Firstly, the detector alignment and the alignment of the grating in the system (see below) is simplified since the active sensing area is larger than the individual pinhole images and the angular range of the dispersion.  Additionally, the fine sampling of the chosen CCD enables accurate centroid determination for verifying the dispersion of each grating, and allows for the use of custom photometric apertures.

\begin{figure}[t]
   \begin{center}
   \begin{tabular}{c}
   \includegraphics[width=\textwidth]{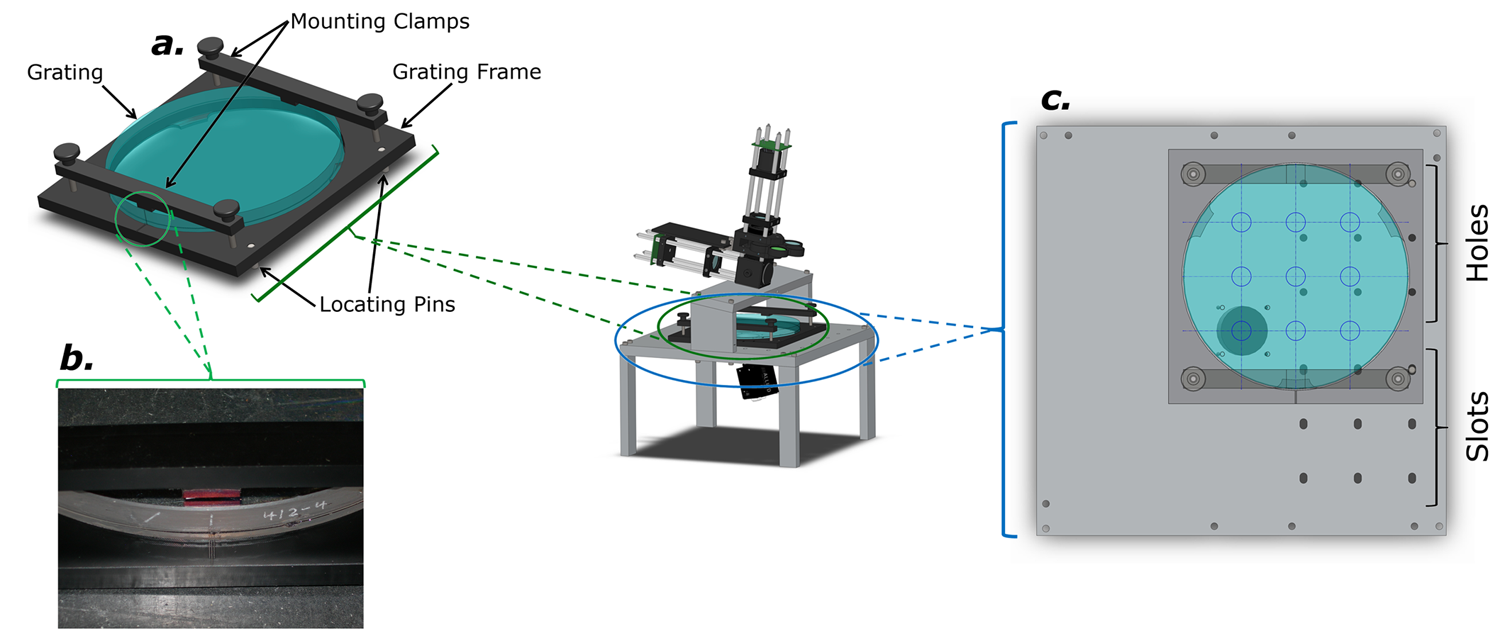}
   \end{tabular}
   \end{center}
   \caption[example] 
   { \label{fig:gratingcell} 
The mechanical design of the VGT grating mounting cell. \textit{a}) A rendering of the grating cell with major components labeled. \textit{b}) The engraving on the edge of one of the prototype VIRUS gratings that indicates the fringe orientation, including the direction of the fringe tilt. This engraving is aligned with a matching feature on the grating cell's frame for rapid initial alignment of the grating. \textit{c}) A rendering of the VGT base showing the series of holes and slots used for maintaining the rotational alignment of the grating when switching between the 9 subapertures. The positions of the subapertures are indicated by the thin blue circles on the grating face.
}
   \end{figure}

Due to imperfections in the grating fabrication process, the diffraction efficiency of a grating may be spatially variable across the CA\cite{Chonis12}. Ideally, one would measure the diffraction efficiency with a collimated beam matched to the CA of the grating. However such an apparatus for our 138 mm diameter gratings would be too large to be portable and would significantly increase costs. To maximize the tested area and provide an empirical estimate of the spatial variability of the diffraction efficiency with our small optical system, we have designed the VGT to easily measure up to 9 subapertures of a grating. Fig. \ref{fig:gratingcell} details the design of a special mounting cell that makes these multiple measurements possible. A marking is placed on the edge of each grating by the vendor to indicate the fringe tilt and fringe direction within $\pm1$\degree\ so that the grating can easily be placed in the VGT mounting cell in the correct orientation. A corresponding mark on the edge of the mounting cell is used for visual alignment of the grating and ensures that the placement of the focused spot is on the CCD chip. The cell contains two press-fit locating pins which mate to a series of holes and slots on the custom tester base for constraining the rotational alignment of the grating as it is moved between the series of 9 test positions.

%-------------
%------------- 

The optomechanical design of the tester has been simplified by utilizing off-the-shelf components for 1'' diameter optics, and modifications to these stock parts were made where necessary to meet our needs. Custom aluminum hardware seamlessly mates with these stock components to fix the collimator and camera angles in the same configuration as a VIRUS spectrograph to an accuracy of $\pm0.1$\degree. Including its light-tight enclosure that allows accurate testing in a fully-illuminated room, the VGT is compact and can easily fit on an office desk. It has the approximate dimensions of 485 mm tall with a 350$\times$320 mm footprint and weighs $\sim12$ kg. 

The LEDs and the comparison photodiode are controlled through a data acquisition unit requiring a single USB connection to a host computer. The CCD camera interfaces with the computer through a Gigabit ethernet port and is powered by an external 12 V DC source. Both the CCD and the comparison photodiode were verified for linearity over the relevant signal levels. The operation of the VGT is controlled through custom Python software that provides near ``push-button'' simplicity for the tester's operation in a command shell environment, and includes the automated reduction of the CCD images, photometry, background subtraction of the photodiode signal, and calculation of diffraction efficiencies. Before shipment to the vendor, an absolute calibration of the VGT was provided by assembling the camera lens and CCD in a ``straight-through'' configuration without the grating and performing CCD photometry on the direct pinhole images of the LEDs. This initial calibration is then refined as needed over time automatically through the continuous monitoring of the LED output by the comparison photodiode, and manually through regular checks of a standardized 930 line mm$^{-1}$ reference grating whose absolute external diffraction efficiency is well-known through measurements made with the flexible test-bench facility discussed in $\S$\ref{subsec:prototypegratings}. The statistical uncertainties in the diffraction efficiency measurements by the VGT are $\pm0.8$\%, $\pm0.8$\%, and $\pm0.1$\% at 350, 450, and 550 nm, respectively.

\subsection{Measurements}\label{subsec:measurements}
%-------------
   \begin{figure}[t]
   \begin{center}
   \begin{tabular}{c}
   \includegraphics[width=0.9\textwidth]{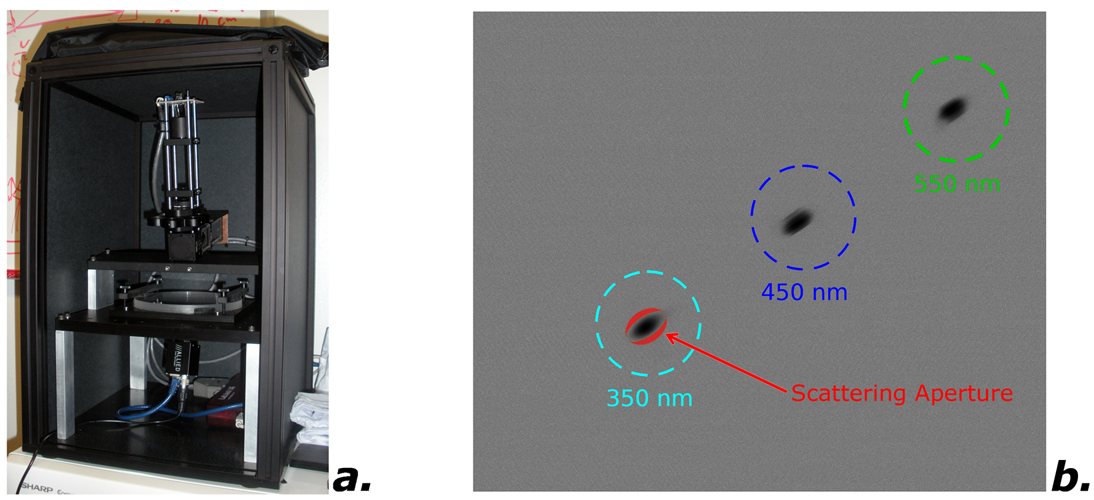}
   \end{tabular}
   \end{center}
   \caption[example] 
   { \label{fig:constructed} 
\textit{a}) A photo of the completed VGT. When in use, a curtain drops over the open face of the enclosure to make the apparatus light-tight. \textit{b}) An image constructed from the sum of three dark subtracted CCD images, each taken with one of three LEDs turned on. Each CCD image was normalized to the peak flux within each indicated circular aperture before the summation. The three circular apertures are used for the photometric measurements from which the diffraction efficiency at each wavelength is calculated. The red elliptical annulus around the 350 nm pinhole image has a maximum angular width of 0.5\degree and was used to measure the near-UV scattering of the gratings. 
}
   \end{figure} 
%------------- 
A photo of the completed VGT can be seen in panel $a$ of Fig. \ref{fig:constructed}. As we demonstrated in Ref. \citenum{Chonis12}, a full test of a grating requires $\lesssim10$ minutes to complete. Thanks to the fixed and rugged nature of the VGT design, the grating acceptance measurements are tailored to verifying the most critical specifications and have become standardized to avoid the unnecessary confusion that results from inconsistent measurement methods. For each VPH grating on the production line, the VGT provides the following key measurements:

\begin{itemize} \itemsep1pt \parskip0pt \parsep0pt
	\item \textbf{Average External Diffraction Efficiency:} Each grating is tested at 350, 450, and 550 nm in each of the 9 different subaperture positions across the grating CA. With a 12.5 mm diameter beam size, this allows 7.4\% of the total CA area to be directly tested. The reported external diffraction efficiency for each wavelength that is to be compared with the specification in Table \ref{tab:Efficiency} is calculated as the average over these 9 subapertures. Fig. \ref{fig:constructed}$b$ shows an example of the pinhole images for each wavelength on the VGT CCD and the location of the over-sized apertures used for performing the photometry.
	\item \textbf{External Diffraction Efficiency Spatial Uniformity:} Using the data gathered above for the 9 subapertures, an estimate of the spatial uniformity of the external diffraction efficiency can be calculated. For example, a simple metric for estimating the spatial diffraction efficiency uniformity is the difference between the maximum and minimum measured efficiencies at each wavelength.
	\item \textbf{Near-UV Scattered Light:} Scattered light within the processed grating layer can be the result of aberrations, reflections, and other imperfections in the optical system used to expose the holographic material in which the grating is formed\cite{Barden01}. Additional sources of scattering are the epoxy layer used to bond the substrates and the surface roughness of the substrates themselves. An example of a grating with particularly bad scattering properties can be found in Fig. 18 of Ref. \citenum{Barden01}. In VIRUS, a grating that scatters a high amount of the incident beam can adversely affect the image quality of the spectrograph leading to an increase in the cross-talk between imaged fibers at the focal plane. Scattering is most pronounced at short wavelengths, so we quantify the worst-case scenario by making the measurement at 350 nm. To increase the signal-to-noise ratio for measuring faint scattered light, the images taken at all 9 subaperture positions are coadded. Two custom photometric apertures are used. First, an inner elliptical aperture is centered on the 350 nm pinhole image with major and minor axes that correspond to the ideal image size as modeled with Zemax, given the dispersion resulting from the non-monochromatic LED output. A second elliptical aperture is also used with the same center and major axis as the first, but with a minor axis that extends an additional angular distance of 0.5\degree\ on either side of the inner elliptical aperture. The total flux within the large circular aperture used in the 350 nm diffraction efficiency calculation is first measured, followed by a measurement of the flux contained within the elliptical annulus formed between the two elliptical apertures (see Fig. \ref{fig:constructed}$b$). The goal for science-grade gratings is to have $\lesssim3$\% of the total flux scattered into the elliptical annulus for the grating to be accepted\footnote{The original specification for scattered light by the gratings stated that $\lesssim3$\% of light at $\lambda=350$ nm in a point source can be scattered into a 0.5\degree\ solid angle cone around the $m=1$ beam at the design $\alpha$. However, as stated in $\S$\ref{subsec:testerdesign}, the narrowband-filtered LED light sources in the VGT are not monochromatic. As a result, the pinhole images on the CCD are elongated in the dispersion direction, which motivates the use of an elliptical annulus as the scattering aperture.}.
\end{itemize}

The most notable optical property that the VGT does not verify is the transmitted wavefront error (TWE). TWE measurements are made on selected gratings to verify consistency from batch to batch using a Zygo interferometer with a 6'' diameter beam (stopped down to a 138 mm diameter) at $\lambda=632$ nm in a double-pass configuration using a reference mirror. The TWE for a science-grade grating should be $<2$ waves peak-to-valley at 632 nm within the CA, including any spherical wavefront error.
%%%%%%%%%%%%%%%%%%%%%%%%%%%%%%%%%%%%%%%%%%%%%%%%%%%%

%%%%%%%%%%%%%%%%%%%%%%%%%%%%%%%%%%%%%%%%%%%%%%%%%%%%
\section{MASS PRODUCTION OF VPH GRATINGS} \label{sec:production}
In this section, the production line process that was used to fabricate the VIRUS VPH gratings at Syzygy Optics, LLC is summarized. For an individual VPH grating, the process begins by preparing a solution of ammonium dichromate and gelatin (i.e., dichromated gelatin; DCG) that serves as the holographic medium. This solution is poured between a glass mold and the base fused silica substrate. To achieve a layer of gelatin that is initially $\sim100$ $\mu$m thick, we attached adhesives to the substrates around the perimeter to serve as shims between the substrate and mold. We then cooled the gelatin until it congealed and removed the substrate from the mold using a releasing agent. The substrate is then dried and subsequently cured in an incubator.

To form a holographic image in the gelatin layer, each cured substrate is exposed to a 457.5 nm coherent light source that has interference in a plane that forms a 1\degree\ angle with the gelatin layer. After the exposure, the grating is submerged in photographic fixer and then dehydrated with graded alcohol. The amount of time in the fixer and in each of the alcohol baths can be varied to produce different modulations of the gelatin layer's index of refraction. Upon removal from the final alcohol bath, we dried any remaining liquid from the grating by placing it in an oven for approximately five minutes. To obtain a preliminary analysis of the optical properties of the uncapped grating\footnote{On average, the exposed and processed DCG layer has approximately the same index of refraction as the glass that is used to cap the grating\cite{Barden00}. As a result, the VGT with its fixed 9\degree\ angle of incidence can be used to measure a grating with or without the cap substrate in place. The only difference in the measurement without the cap substrate is the lack of an AR coated incident surface, and the lack of losses due to the internal transmittance of the fused silica. Both of these effects can be accounted for to estimate the final diffraction efficiency of the capped grating.}, we took a series of rapid measurements with the VGT to determine if the grating met the minimum required efficiency at 350, 450, and 550 nm. If the grating did not meet the minimum diffraction efficiency specification, the process of fixing and dehydration is immediately modified for the following grating. If necessary, gratings can be reprocessed through the alcohol baths to further modulate the gelatin layer's refractive index. If a grating was not uniform across the CA, however, it would not be eligible for reprocessing. Using the VGT to determine the level of uniformity in-situ prevented the reprocessing of gratings that could not consistently reach the minimum required diffraction efficiency across the CA.

Acceptable gratings were stored in a dry box for 2-3 days before retesting with the VGT to confirm that the grating properties did not significantly vary from the initial measurements. If there were no significant changes after drying, a 4-5 mm ring of the gelatin around the edge of the grating is removed and the grating is capped with the second fused silica substrate using an optical grade glue. The ring around the edge of the grating that is devoid of gelatin fills with the adhesive and encapsulates the diffractive medium. This seals the grating and prevents moisture from entering and altering the material. Once the adhesive sets, the VGT is then used for the final measurements at each of the 9 subapertures before being approved as science-grade. In the following section, we present the final VGT measurements for the 170 science-grade gratings that were fabricated using this production process.
%%%%%%%%%%%%%%%%%%%%%%%%%%%%%%%%%%%%%%%%%%%%%%%%%%%%

%%%%%%%%%%%%%%%%%%%%%%%%%%%%%%%%%%%%%%%%%%%%%%%%%%%%
\section{PERFORMANCE OF THE VPH GRATING SUITE} \label{sec:performance}
To be accepted as a science-grade grating, each of the 170 units must meet the basic assembly specifications. These include having a total thickness of $16.0\pm0.5$ mm, a physical diameter of $148.0\pm0.5$ mm, a radial mismatch of the two fused-silica substrates of $<0.5$ mm, and a total wedge of $<30$\arcmin\ and $<10$\arcmin\ perpendicular and parallel to the fringes, respectively. The VPH layer must also have a CA with a diameter $>138.0$ mm, be centered on the base substrate to within $\pm1$ mm, and be free of major bubbles and point defects. Averaged over all 170 science-grade gratings, we have measured the mean defect area to be 1.13 mm$^{2}$ (standard deviation $\sigma = 0.53$ mm$^{2}$). The maximum defect area of any individual science-grade grating is 2.40 mm$^{2}$, which corresponds to only 2.0\% of the area of a single VGT subaperture. As a result, our characterization of the average external diffraction efficiency will not be significantly affected by measuring a subaperture that contains a large bubble or point defect. Each science-grade grating must have no chips within the CA on either substrate, be able to meet a surface finish specification of 60/40 scratch/dig, and have a surface roughness of $<2$ nm within the CA. Finally, each VPH grating assembly was supplied with a unique serial number that can be traced back to a specific manufacturing date and process during production. 

\subsection{Average External Diffraction Efficiency}\label{subsec:efficiency}
%-------------
   \begin{figure}[t]
   \begin{center}
   \begin{tabular}{c}
   \includegraphics[width=0.99\textwidth]{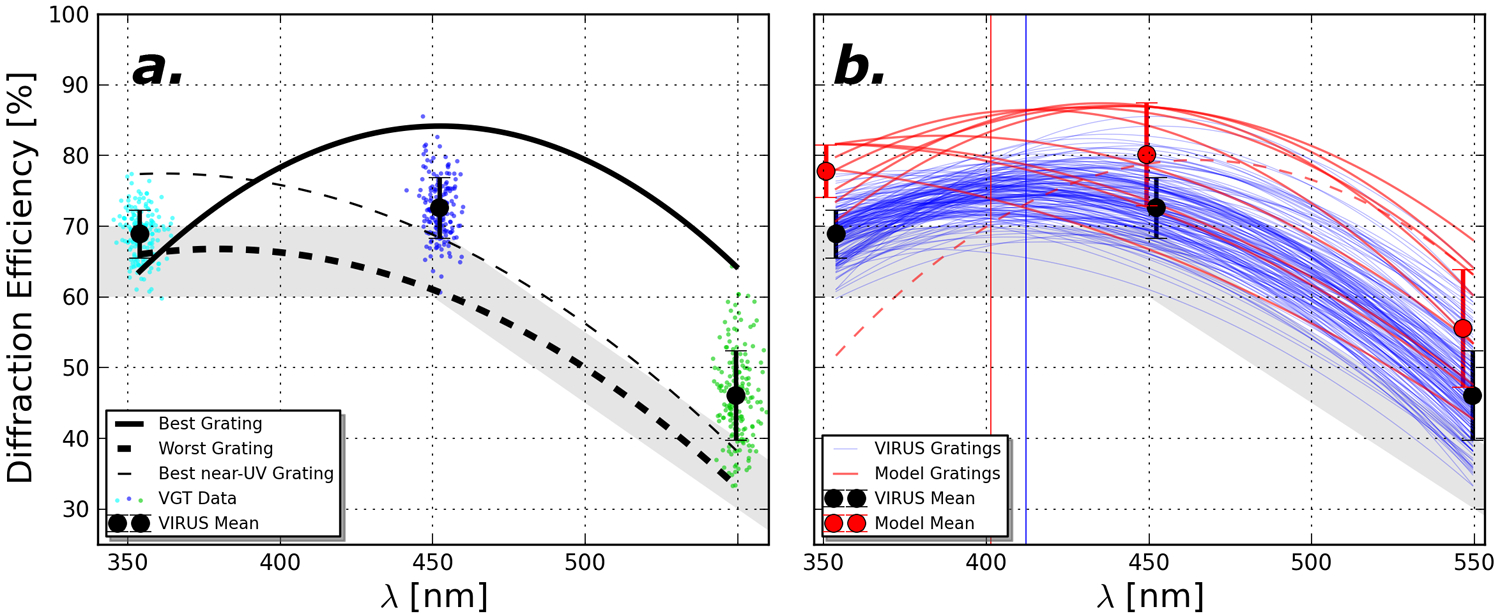}
   \end{tabular}
   \end{center}
   \caption[example] 
   { \label{fig:Results} 
The average delivered external diffraction efficiency of the VIRUS VPH grating suite. The shaded gray region in both panels indicates the external diffraction efficiency specification outlined in Table \ref{tab:Efficiency}. \textit{a}) The average VGT diffraction efficiency measurements for each grating are shown as the small colored data points (teal for 350 nm, blue for 450 nm, and green for 550 nm), which have been scattered randomly about the tested wavelengths for visual clarity. The large black data points show the average external diffraction efficiency of the entire grating suite at each of the three measured wavelengths, while the error bars show the standard deviation of the distributions. The black curves show quadratic spline fits to the diffraction efficiency measurements for notable individual gratings (the heavy solid and heavy dashed curves correspond to the best and worst overall gratings, respectively, while the light dashed curve is the grating that performs best at 350 nm with a peak external diffraction efficiency of 77.4\%). \textit{b}) A comparison of the delivered grating suite to 10 gratings simulated with RCWA. The blue curves are quadratic spline fits to the individual VGT grating measurements, while the red curves are quadratic spline fits to the simulated gratings. The large black (red) data points and error bars are the mean diffraction efficiency and standard deviation for the delivered (simulated) grating suites. The dashed red curve indicates a simulated grating that would be rejected as science-grade. The vertical blue (red) lines indicate the average wavelength of peak diffraction efficiency for the delivered (simulated) grating suites. See the text of $\S$\ref{subsec:efficiency} for more details.  
}
   \end{figure} 
%------------- 
%-------------
   \begin{figure}[t]
   \begin{center}
   \begin{tabular}{c}
   \includegraphics[width=0.99\textwidth]{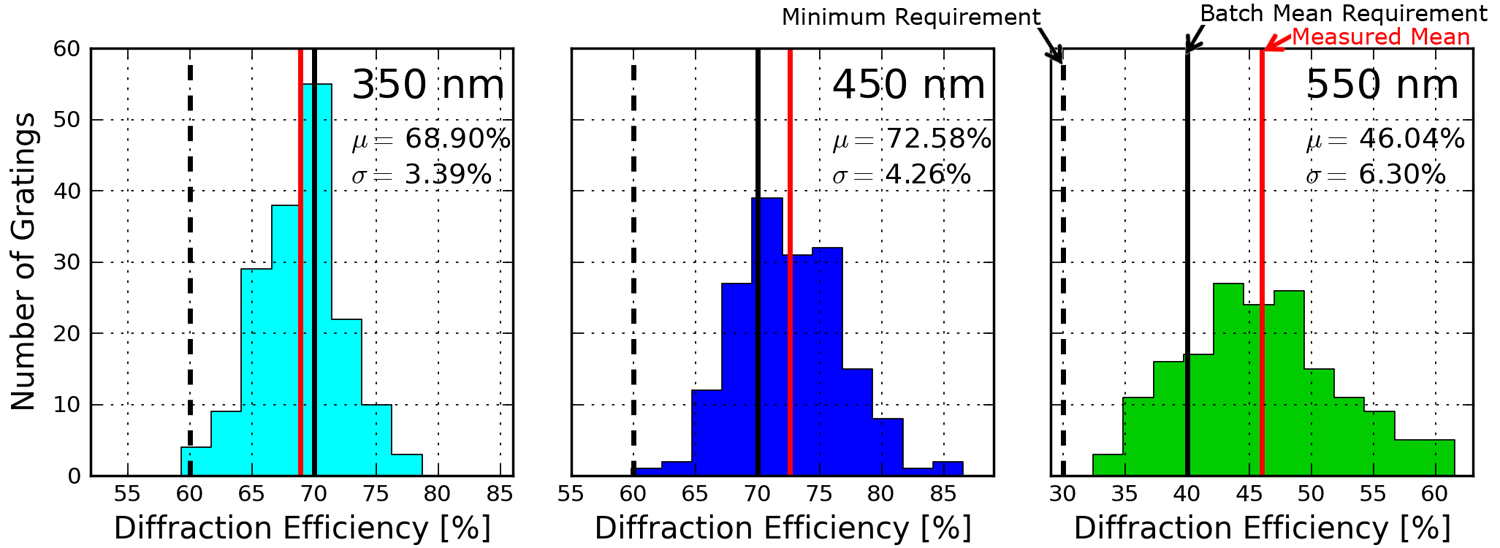}
   \end{tabular}
   \end{center}
   \caption[example] 
   { \label{fig:EfficiencyHist} 
Histograms showing the distribution of the external diffraction efficiency for the delivered VIRUS VPH grating suite, as measured by the VGT. Each histogram represents a cross-section cut along the $y$-axis at each wavelength from Figure \ref{fig:Results}$a$. For each panel, the solid and dashed black vertical line indicates the batch mean and minimum requirements on the external diffraction efficiency, respectively, while the solid red vertical line indicates the measured mean external diffraction efficiency for the entire grating suite.\vspace{5mm}
}
   \end{figure} 
%------------- 

The external diffraction efficiency results for the delivered science-grade VPH gratings for VIRUS are shown in Fig. \ref{fig:Results}$a$. For each of the three tested wavelengths, we show the individual, spatially averaged VGT diffraction efficiency measurements along with the suite mean and standard deviation averaged over all 170 gratings. In addition, the distribution of the diffraction efficiency for each wavelength is shown in Fig. \ref{fig:EfficiencyHist}. The histograms in that figure represent a cross-section at each wavelength along the $y$-axis of Fig. \ref{fig:Results}$a$. These results are also summarized in Table \ref{tab:Results}. In general, the measured external diffraction efficiency of the VPH grating suite very closely meets or exceeds our batch mean requirement at each of the three tested wavelengths on average. Additionally, the majority of the delivered gratings greatly exceed the minimum external diffraction efficiency specification, with only a single grating dropping below the minimum requirement at 350 nm by 0.2\%. 

To quickly compare the individual gratings amongst each other quantitatively, we calculate $\int\!\eta(\lambda)\:d\lambda$ for each grating between $350 < \lambda \mathrm{(nm)} < 550$, where $\eta(\lambda)$ is the function describing the external diffraction efficiency curve. Since we only have measurements at three discrete wavelengths, we interpolate the diffraction efficiency between the measurements using a quadratic spline to determine $\eta(\lambda)$ for each grating. In Fig. \ref{fig:Results}$a$, the heavy solid curve shows the quadratic spline fit to the measurements of the best overall performing grating according to this merit function while the heavy dashed curve represents the worst grating. Despite the large difference in performance of these two gratings at longer wavelengths, both have similar efficiency at 350 nm. This is consistent with the measurements in general (see Table \ref{tab:Results}), as will also be seen in the following subsection on the spatial uniformity of the diffraction efficiency. This consistency is the result of communication with the contractor that high efficiency at 350 nm was the primary goal for the grating performance due to the multiple combined effects in the near-UV that make detecting LAEs difficult (e.g., see $\S$\ref{sec:intro}). Of all the gratings in the suite, the best-performing unit at 350 nm has a spatially averaged external diffraction efficiency of 77.4\%. The quadratic spline fit to this unit's VGT data has been highlighted in Fig. \ref{fig:Results}$a$ in addition to that of the best and worst performing gratings mentioned above. Overall, the production process was very consistent with time, as there is no statistical dependence of the external diffraction efficiency at any wavelength on the date of manufacture.

The range of variation from grating to grating seen in Figs. \ref{fig:Results} and \ref{fig:EfficiencyHist} is due to small differences in the properties of the processed DCG layer in which the grating is formed. A VPH grating can be fully described and its diffraction efficiency modeled at a given $\alpha$ and $m$ (e.g., with a Rigorous Coupled Wave Analysis; RCWA\cite{Gaylord85}) given the following properties: the fringe density, the fringe tilt $\phi$, the DCG layer refractive index $n_{\mathrm{DCG}}$ and its sinusoidal modulation $\Delta n_{\mathrm{DCG}}$, and the DCG layer thickness $d$. Of these properties at a fixed $\alpha$ and $m$, those that significantly affect the diffraction efficiency of the grating are $d$, $\Delta n_{\mathrm{DCG}}$, and $\phi$. To estimate the range of variation in these parameters that match the measured unit-to-unit diffraction efficiency variation in the delivered grating suite, we ran a series of $m=1$ RCWA models based around the targeted parameters for VIRUS. As described in Ref. \citenum{Chonis12}, those parameters are 930 line mm$^{-1}$ fringe density, $\phi=-1$\degree, $\alpha=9$\degree, $d=5.5$ $\mu$m, and $\Delta n_{\mathrm{DCG}}=0.037$. We assume that $n_{\mathrm{DCG}} = 1.5$ (e.g., Ref. \citenum{Barden00}), and apply factors to the RCWA modeled diffraction efficiency to take into account the transmission through the AR coated fused-silica substrates, the epoxy layer, and the transmittance of the DCG layer itself for typical physical thicknesses\cite{Barden00}. The predicted diffraction efficiency was calculated at 350, 450 and 550 nm to mimic the VGT measurements. In Fig. \ref{fig:Results}$b$, we show a sample suite of 10 RCWA modeled gratings (red quadratic splines) compared to the delivered VIRUS grating suite (blue quadratic splines). The exact values of $d$, $\Delta n_{\mathrm{DCG}}$, and $\phi$ for each RCWA model were chosen from a uniform distribution with a range about the targeted value of $\pm1.0$ $\mu$m, $\pm0.01$, and $\pm0.5$\degree, respectively. As can be seen, the range of diffraction efficiency at each measured wavelength in the RCWA model suite qualitatively matches that of the delivered grating suite relatively well. Similar to what we observe for the delivered gratings, the wavelength with the least variation from unit to unit is 350 nm. Additionally, the overall shape of the average modeled diffraction efficiency curve matches that of the delivered grating suite well, with a difference in the average diffraction efficiency peak of only 10.8 nm (see the vertical lines in Fig. \ref{fig:Results}$b$). Of the modeled gratings, one was not considered in the discussion above due to not meeting the diffraction efficiency requirements for a science-grade grating (see the dashed red curve; the fall-off at 350 nm was due to an extremely large $\Delta n_{\mathrm{DCG}}$ coupled with $d$ that was also larger than the targeted value). In general, however, the average diffraction efficiency of the delivered gratings falls systematically short of the RCWA models at each wavelength (also, see Ref. \citenum{Chonis12} and $\S$\ref{subsec:uniformity} below). The exercise outlined above gives an estimation of the precision to which the DCG layer can be processed for modern VPH gratings.    

\begin{table}[t]
\caption{Summary of the VGT performance measurements averaged over the VIRUS VPH diffraction grating suite. The standard deviation among the sample of 170 gratings for each quantity is shown in parentheses. The wavelengths listed are the effective VGT measurement wavelengths (i.e., the wavelength centroids of the narrowband-filtered LED spectra). The ``Diffraction Efficiency'' measurements are described in $\S$\ref{subsec:efficiency}, the various ``Spatial Variation'' measures are described in $\S$\ref{subsec:uniformity}, and the ``Scattering'' measurement are described in $\S$\ref{subsec:scattering}.} 
\label{tab:Results}
\begin{center}       
\begin{tabular}{|c|c|c|c|c|c|} 
\hline
\rule[-1ex]{0pt}{3.5ex}  \textbf{$\lambda$} & \textbf{Diffraction Efficiency} & \textbf{Spatial Variation} & \textbf{} & \textbf{} & \textbf{Scattering} \\
\rule[-2ex]{0pt}{1ex}  \textbf{[nm]} & \textbf{[\%]} & \textbf{Total [\%]} & \textbf{High [\%]} & \textbf{Low [\%]} & \textbf{[\%]} \\
\hline
\rule[-1ex]{0pt}{3.5ex}  353.9 & 68.90 (3.39) & 10.76 (5.64) & $+4.21$ (1.92) & $-6.55$ (4.12) & 3.20 (0.37) \\
\hline
\rule[-1ex]{0pt}{3.5ex}  452.3 & 72.58 (4.26) & 19.10 (8.57) & $+7.46$ (3.14) & $-11.63$ (6.11) & $-$ \\
\hline
\rule[-1ex]{0pt}{3.5ex}  549.5 & 46.04 (6.30)& 18.74 (7.13) & $+8.28$ (3.29) & $-10.47$ (4.57) & $-$ \\
\hline
\end{tabular}
\end{center}
\end{table}

\subsection{Spatial Uniformity} \label{subsec:uniformity}
\begin{figure}[t]
   \begin{center}
   \begin{tabular}{c}
   \includegraphics[width=0.99\textwidth]{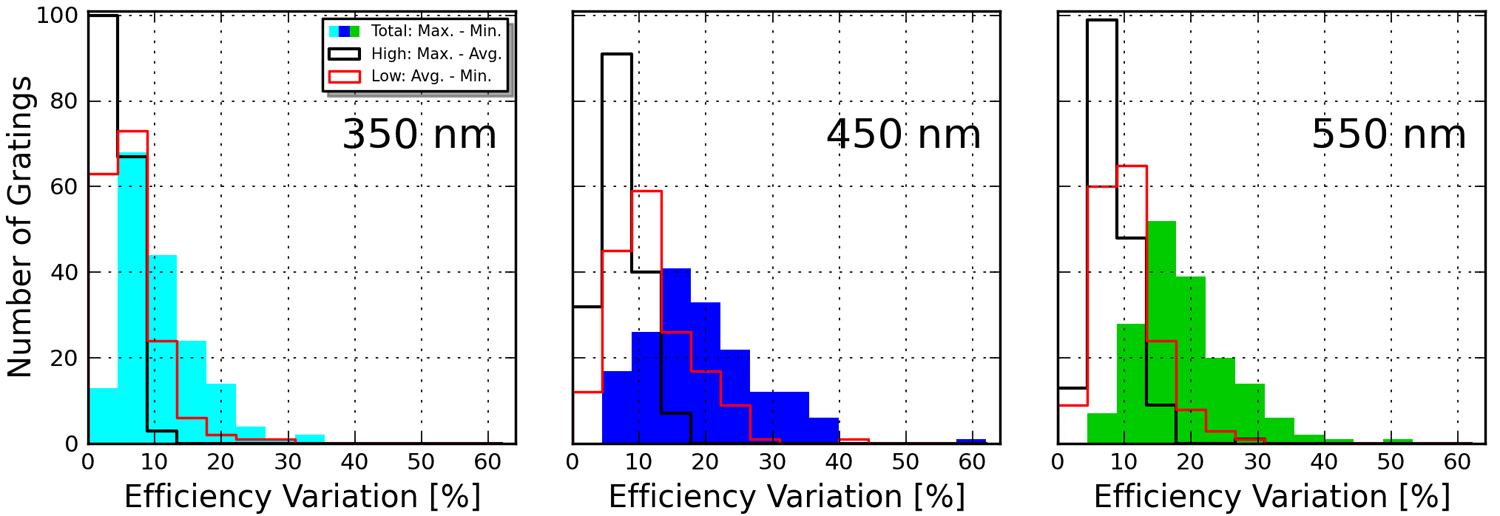}
   \end{tabular}
   \end{center}
   \caption[example] 
   { \label{fig:UniformityHist} 
Histograms showing the distribution of different measures of the uniformity of the external diffraction efficiency for the delivered VIRUS VPH grating suite, as measured by the VGT. Each panel corresponds to one of the three measured wavelengths and contains three histograms. The shaded histograms represent the difference between the maximum and minimum diffraction efficiency measured over the 9 VGT subapertures (``Total''). The black histograms represent the difference between the maximum diffraction efficiency measured over the 9 subapertures and the average (``High''). Finally, the red histograms represent the difference between the minimum diffraction efficiency measured over the 9 subapertures and the average (``Low''). To ease the comparison of the ``High'' and ``Low'' distributions, the ``Low'' distribution (which consists entirely of negative values) is shown as positive.
}
   \end{figure}

From the measurement of the 9 individual subapertures across a grating CA, the VGT provides a measure of the spatial uniformity of the external diffraction efficiency. As labeled in Table \ref{tab:Results}, the ``Total'' spatial variation is a simple measure of uniformity calculated by taking the difference between the maximum and the minimum measured diffraction efficiencies between the 9 subapertures. In Fig. \ref{fig:UniformityHist}, we show the distributions of the total spatial variation for the VIRUS VPH grating suite at each of the three measured wavelengths. As was the case with the average external diffraction efficiency discussed in the previous subsection, the most consistent performance is at 350 nm, where the mean total spatial variation is $\sim2\times$ smaller than at 450 or 550 nm. The 350 nm distribution's standard deviation is also significantly smaller than the other two wavelengths.  

%-------------
%-------------
%-------------
   \begin{figure}[ht]
   \begin{center}
   \begin{tabular}{c}
   \includegraphics[width=0.99\textwidth]{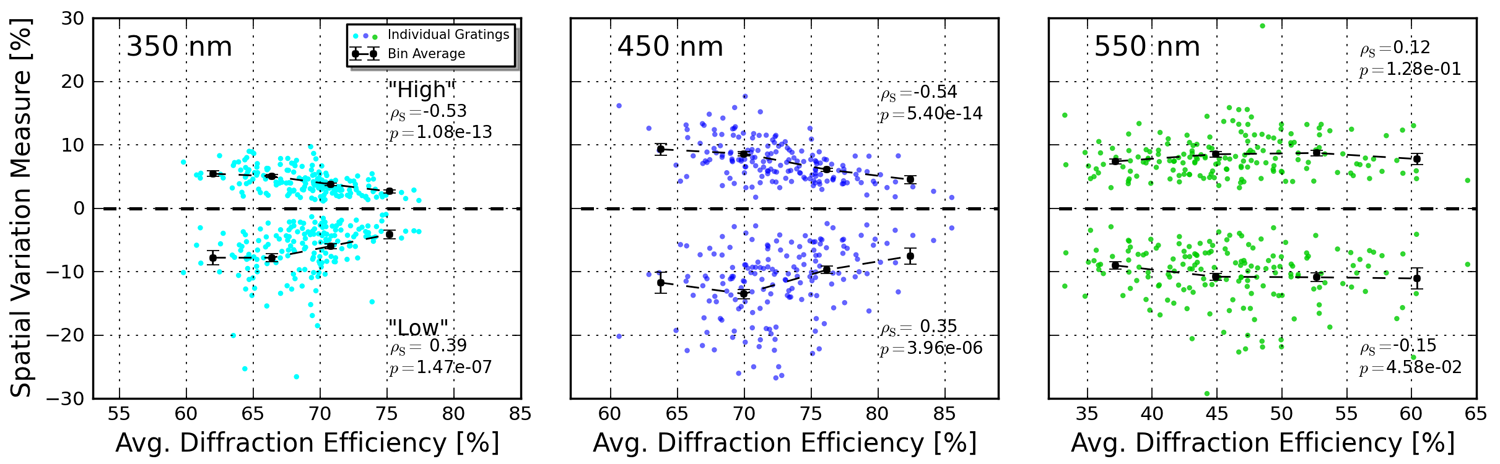}
   \end{tabular}
   \end{center}
   \caption[example] 
   { \label{fig:UniformityCorrelation} 
Scatter plots of the ``High'' and ``Low'' spatial variation measures as a function of the spatially averaged diffraction efficiency for each VPH grating in the VIRUS suite at each of the three wavelengths measured by the VGT. As described in $\S$\ref{subsec:uniformity}, the ``High'' (``Low'') spatial variation is simply the difference between the maximum (minimum) diffraction efficiency measured over the 9 subapertures and the spatially averaged diffraction efficiency that is plotted on the abscissa. Thus, there are two colored data points plotted for each grating: one above the thick dashed line at $y=0$, and one below. The larger black data points in each panel correspond to the average spatial variation measure calculated for four equally spaced bins of average diffraction efficiency, while the error bars correspond to the standard error of the mean.
}
   \end{figure} 
%-------------

In addition to the total spatial variation, we also look at the spatial variation above and below the average external diffraction efficiency. As labeled in Table \ref{tab:Results}, the ``High'' spatial variation is simply the difference between the maximum measured diffraction efficiency among the 9 subapertures and the spatially averaged diffraction efficiency for a given grating. Similarly, the ``Low'' spatial variation is the difference between the minimum measured diffraction efficiency among the 9 subapertures and the spatially averaged diffraction efficiency, and is a negative quantity. The ``High'' (``Low'') distributions for the VIRUS VPH grating suite are also shown in the panels of Fig. \ref{fig:UniformityHist} as the black (red) histograms. In Fig. \ref{fig:UniformityHist}, the ``Low'' distributions are shown as positive to facilitate visual comparison with the ``High'' distributions. By simply looking at the mean and standard deviation of the two distributions for each wavelength (see Table \ref{tab:Results}), it is clear that the distributions significantly differ. This is confirmed by running a two-sample Kolmogorov-Smirnov test on the ``High'' and ``Low'' distributions, which indicates that the null-hypotheses (i.e., the two distributions are drawn from the same parent distribution) can be rejected at each wavelength with very high certainty. At each wavelength, the ``High'' distribution is narrower and has a lower mean than the corresponding ``Low'' distribution. This is likely due to the fact that the maximal diffraction efficiency at a given wavelength for a fixed fringe density and $\alpha$ is achieved for exactly the right combination of properties that describe the processed DCG layer (i.e., primarily $\Delta n_{\mathrm{DCG}}$ and $d$)\cite{Barden00,Baldry04}. Given some distribution of achieved values for these processed DCG layer properties about the targeted values, it is more likely that a combination of non-optimal values will be drawn rather than drawing the exact correct set of values that maximizes the diffraction efficiency. As a result, non-uniformities in the DCG layer processing that cause the diffraction efficiency to vary across the grating CA (e.g., small changes in effective gelatin thickness) are more likely to reduce the diffraction efficiency than boost it about the average in a given subaperture. The result is that the ``High'' distribution is narrower and smaller on average than that for the ``Low'' distribution. Since the spatial variations in DCG layer properties tend to decrease the measured average diffraction efficiency for a given grating, it should not be surprising that the modeled RCWA predictions discussed in the previous subsection and shown in Fig. \ref{fig:Results}$b$ (which does not consider the spatial variation of grating parameters) are systematically higher than the measurements for the delivered gratings. 

Fig. \ref{fig:UniformityCorrelation} also supports the aforementioned hypotheses that small changes in the DCG properties as a function of position across the CA tend to scatter individual measurements towards lower efficiency rather than higher efficiency. In this figure, we plot both the ``High'' and ``Low'' spatial variation measures for each grating as a function of the spatially averaged diffraction efficiency for each respective wavelength. In addition, we have calculated the Spearman Rank Correlation Coefficient $\rho_{\mathrm{S}}$ and associated $p$-value for the ``High'' and ``Low'' variation measures separately at each wavelength. As can be seen in Fig. \ref{fig:UniformityCorrelation}, the ``High'' and ``Low'' variation measures are correlated with high statistical significance at 350 and 450 nm such that more uniform gratings have higher spatially averaged diffraction efficiency. However, this trend is not seen at 550 nm. For 350 and 450 nm, the fact that the ``Low'' spatial variation measure increases with increasing average diffraction efficiency is not surprising. What is more interesting is that the ``High'' spatial variation measure \textit{decreases} with increasing average diffraction efficiency. This is likely a result of the highest average diffraction efficiency gratings already having close to the optimal DCG properties that maximize the diffraction efficiency. As a result, non-uniformities in the DCG layer processing are increasingly less likely to result in a better combination of layer properties. The reason these trends are not seen at 550 nm is because it is the furthest tested wavelength from the Bragg condition (from $\S$\ref{sec:gratingspec}, recall that the VIRUS gratings were designed such that the  Bragg wavelength is between $350 < \lambda\; \mathrm{(nm)} < 400$; from Fig. \ref{fig:Results}$a$, the average wavelength of the peak diffraction efficiency for the suite is 412.3 nm). As a result, the diffraction efficiency at 550 nm is not maximal, yielding a more equal likelihood that a slight change in DCG properties could either scatter the diffraction efficiency positively or negatively about the average. 

\subsection{Near-UV Scattering}\label{subsec:scattering}
\begin{figure}[t]
   \begin{center}
   \begin{tabular}{c}
   \includegraphics[width=0.7\textwidth]{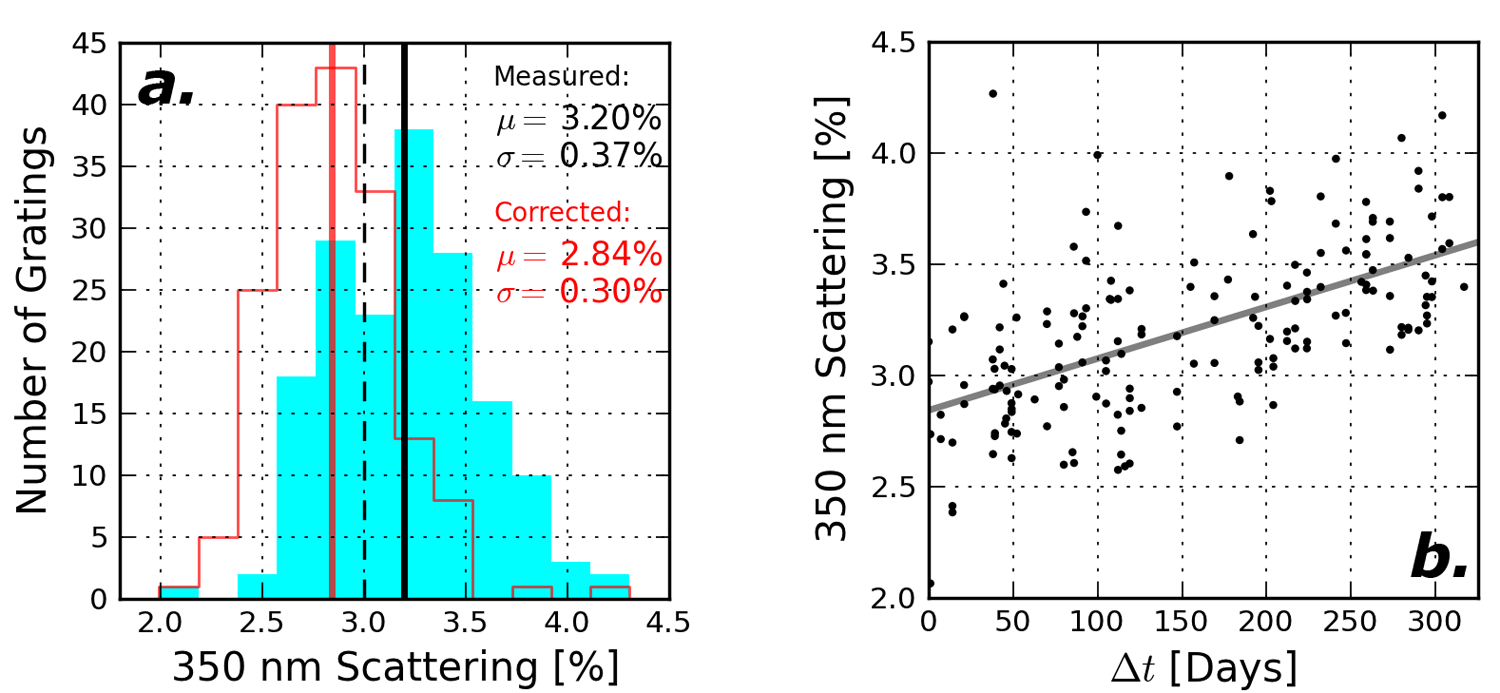}
   \end{tabular}
   \end{center}
   \caption[example] 
   { \label{fig:Scattering} 
The scattered light properties of the VIRUS VPH grating suite. $a$) The distribution of the scattered light at 350 nm as measured by the VGT. The teal histogram represents the original measurements, while the red histogram represents the measurements after correction for the scattering time dependence that is suspected to be due to an increasing amount of dust on the VGT optics (see panel $b$). The dashed black vertical line indicates our original specification on the fraction of the 350 nm pinhole image's total flux that could be scattered into the VGT scattering aperture ($\lesssim3$\%; see $\S$\ref{subsec:measurements}). The solid black vertical line indicates the mean of the entire VIRUS VPH grating suite from the original measurements, and the solid red vertical line indicates the mean after the correction. $b$) The measured scattered light at 350 nm as a function of manufacturing date. The manufacturing date is represented as $\Delta t$ in days, which is the time since the completion of the first science-grade grating. The gray line represents a linear fit to the data and is used as an estimation of the increased scattering effect due to the increasing dust deposited on the VGT optics with time. The slope of this linear fit is used to correct the original measurements, and the corrected data are shown as the red histogram in panel $a$.
}
   \end{figure}

Fig. \ref{fig:Scattering}$a$ shows the distribution of the scattering measurements made by the VGT at 350 nm. On average, the fraction of light in the 350 nm pinhole image that is scattered into the VGT scattering aperture (see $\S$\ref{subsec:measurements}) is 3.2\%, which is slightly worse than our desired $\lesssim3$\% specification. With a standard deviation of only 0.37\%, all gratings in the VIRUS VPH grating suite perform comparably, and the amount of observed scattering in the worst performing grating (4.3\%) should not significantly increase the fiber-to-fiber cross-talk on the CCD detector beyond what is acceptable for HETDEX. 

%-------------
%-------------

%-------------
   \begin{figure}[t]
   \begin{center}
   \begin{tabular}{c}
   \includegraphics[width=0.8\textwidth]{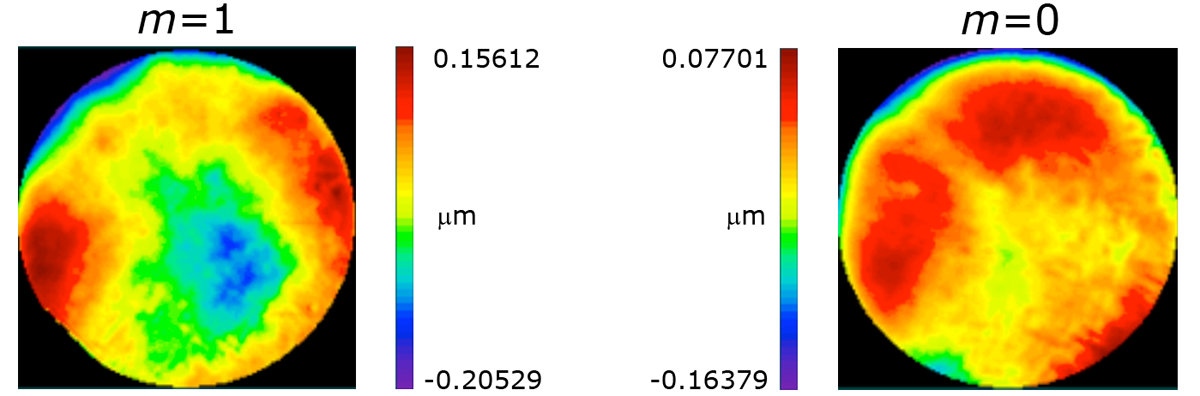}
   \end{tabular}
   \end{center}
   \caption[example] 
   { \label{fig:twe} 
Surface/wavefront maps showing the TWE at $\lambda=632$ nm for a typical grating in the VIRUS VPH grating suite. The map on the left shows the $m=1$ TWE while the map on the right shows the $m=0$ TWE. The typical $m=1$ TWE is $1/2$ wave peak-to-valley at $\lambda=632$ nm, which is significantly better than the $\leq2$ wave peak-to-valley requirement for VIRUS. 
}
   \end{figure} 
%-------------

Of all the properties that were measured by the VGT, the near-UV scattering is the only one that is significantly correlated with the manufacturing date ($\rho_{\mathrm{S}}=0.61$, $p=1.07\times10^{-18}$; see Fig. \ref{fig:Scattering}$b$). Recall that the VGT operates in an office environment rather than a cleanroom. Without clean air in circulation around the VGT during the year over which these tests were carried out, we suspect that the slow increase in the measured scattered light with time is a systematic effect caused by the steady increase of dust deposited on the VGT's lenses and filters, rather than an increase in the actual grating layer scattering. To estimate the average scattering in this scenario, we assume that the VGT optics were clean as of the time of the first science-grade grating's completion, and simply subtract the slope of a linear fit to the scattering data as a function of time. The linear fit can be seen in Fig. \ref{fig:Scattering}$b$ and the resulting histogram of corrected scattering measurements can be found in Fig. \ref{fig:Scattering}$a$. The mean and standard deviation of the corrected distribution is 2.84\% and 0.30\%, respectively. After the correction, 48 gratings ($\sim28$\% of the suite) have a scattering measurement of $>3$\%.

\subsection{Transmitted Wavefront Error}\label{subsec:twe}
Fig. \ref{fig:twe} shows surface/wavefront maps for a typical VIRUS VPH grating as measured at $m=0$ and $m=1$. The surface errors do not appear to be correlated among the two measured spectral orders. The typical TWE at $\lambda=632$ nm for $m=1$ ($m=0$) is $\sim360$ nm ($\sim240$ nm) peak-to-valley. For VIRUS, this $1/2$ wave peak-to-valley performance at $m=1$ is excellent and is significantly better than our $\leq2$ wave peak-to-valley requirement. This is the result of a high performance holographic exposure system in addition to the use of thick substrates to reduce the effect of warping after the cap substrate is finally glued over the processed grating layer.
%%%%%%%%%%%%%%%%%%%%%%%%%%%%%%%%%%%%%%%%%%%%%%%%%%%%

%%%%%%%%%%%%%%%%%%%%%%%%%%%%%%%%%%%%%%%%%%%%%%%%%%%%
\section{SUMMARY} \label{sec:conclusions}
In this paper, we have presented the design of the VPH diffraction gratings that have been mass-produced for use in the new VIRUS array of spectrographs for the HET. The grating design was optimized to have high external diffraction efficiency in the near-UV. This is required for VIRUS to maintain sufficient throughput for the HETDEX survey, which aims to constrain dark energy and measure its evolution from $1.9 < z < 3.5$ using LAEs as tracers of large scale structure. One of the principle challenges involved in the production of the suite of 170 gratings is maintaining consistency in the high performance standard required for HETDEX. With such a large number of units, we are also faced with the challenge of efficiently and consistently validating the performance of each of the 170 gratings to ensure that the best quality units are delivered. To perform these tests, we have developed an apparatus that is very effective at providing robust acceptance test results in which measurements of the average external diffraction efficiency, spatial uniformity of the diffraction efficiency, and near-UV scattered light are provided in $\lesssim10$ minutes per grating. We have tested the suite of 170 science-grade gratings and determined that they individually meet or exceed our specifications. At near-UV wavelengths, the average grating in the suite achieves an external diffraction efficiency of $\sim70$\%. As the first optical astronomical instrument to be replicated on such a large scale, the VIRUS project has provided a useful platform on which the production of large aperture VPH gratings for astronomy can be evaluated in a statistical manner.
%%%%%%%%%%%%%%%%%%%%%%%%%%%%%%%%%%%%%%%%%%%%%%%%%%%%

%%%%%%%%%%%%%%%%%%%%%%%%%%%%%%%%%%%%%%%%%%%%%%%%%%%%%%%%%%%%%
\acknowledgments     %>>>> equivalent to \section*{ACKNOWLEDGMENTS}       
HETDEX is run by the University of Texas at Austin McDonald Observatory and Department of Astronomy with participation from the Ludwig-Maximilians-Universit\"{a}t M\"{u}nchen, Max-Planck-Institut f\"{u}r Extraterrestriche-Physik (MPE), Leibniz-Institut f\"{u}r Astrophysik Potsdam (AIP), Texas A\&M University (TAMU), Pennsylvania State University, Institut f\"{u}r Astrophysik G\"{o}ttingen (IAG), University of Oxford, and Max-Planck-Institut f\"{u}r Astrophysik (MPA).  In addition to Institutional support, HETDEX is funded by the National Science Foundation (grant AST-0926815), the State of Texas, the US Air Force (AFRL FA9451-04-2-0355), the Texas Norman Hackerman Advanced Research Program under grants 003658-0005-2006 and 003658-0295-2007, and generous support from private individuals and foundations.

We thank the staffs of McDonald Observatory, AIP, MPE, TAMU, Oxford University Department of Physics, and IAG for their contributions to the development of VIRUS. We also acknowledge Jim Arns of Kaiser Optical Systems, Inc. for useful discussions during the development phase of the VPH gratings for the Mitchell Spectrograph and VIRUS. T.S.C. acknowledges the support of a National Science Foundation Graduate Research Fellowship. 
%%%%%%%%%%%%%%%%%%%%%%%%%%%%%%%%%%%%%%%%%%%%%%%%%%%%%%%%%%%%%

%%%%% References %%%%%
\bibliography{ms_arXiv}         %>>>> bibliography data in report.bib
\bibliographystyle{spiebib}   %>>>> makes bibtex use spiebib.bst
%%%%%%%%%%%%%%%%%%%%%%

\end{document}